\documentclass[journal]{IEEEtran}
\usepackage{amsmath}
\usepackage{amsfonts}
\usepackage{algorithmic}
\usepackage{array}
\usepackage{flushend}
\usepackage[caption=false,font=normalsize,labelfont=sf,textfont=sf]{subfig}
\usepackage{textcomp}
\usepackage{stfloats}
\usepackage[numbers,sort&compress]{natbib}
\usepackage{url}
\usepackage{color}
\usepackage{verbatim}
\usepackage{graphicx}
\usepackage{multirow} 
\def\BibTeX{{\rm B\kern-.05em{\sc i\kern-.025em b}\kern-.08em
    T\kern-.1667em\lower.7ex\hbox{E}\kern-.125emX}}
\usepackage{balance}

\begin{document}
\title{\textsf{MetaOpera}: A Cross-Metaverse Interoperability Protocol}

\author{Taotao~Li, Changlin~Yang\IEEEmembership{, Member,~IEEE,} Qinglin~Yang\IEEEmembership{, Member,~IEEE,} Siqi~Zhou, Huawei~Huang,\IEEEmembership{~Senior Member,~IEEE,} Zibin~Zheng,\IEEEmembership{~Fellow,~IEEE }
\thanks{Copyright (c) 2015 IEEE. Personal use of this material is permitted. However, permission to use this material for any other purposes must be obtained from the IEEE by sending a request to pubs-permissions@ieee.org.}
 \thanks{Corresponding Author: Huawei~Huang, huanghw28@mail.sysu.edu.cn}
}

\markboth{Journal of \LaTeX\ Class Files,~Vol.~18, No.~9, September~2020}
{How to Use the IEEEtran Templates}

\maketitle

\begin{abstract}

With the rapid evolution of metaverse technologies, numerous metaverse applications have arisen for various purposes and scenarios. This makes interoperability across metaverses becomes one of the fundamental technology enablers in the metaverse space. The aim of interoperability is to provide a seamless experience for users to interact with metaverses.
However, the development of cross-metaverse interoperability is still in its initial stage in both industry and academia.
In this paper, we review the state-of-the-art cross-metaverse interoperability schemes. These schemes are designed for specific interoperating scenarios and do not generalize for all types of metaverses. 
%
%
To this end, we propose \textsf{MetaOpera}, a generalized cross-metaverse interoperability protocol. By connecting to the \textsf{MetaOpera}, users and objects in metaverses that rely on centralized servers or decentralized blockchains are able to interoperate with each other.
%
We also develop a proof-of-concept implementation for \textsf{MetaOpera}, evaluate its performance, and compare it with a state-of-the-art cross-metaverse scheme based on Sidechains. Simulation results demonstrate that the size of cross-metaverse proof and the average time of cross-metaverse transactions using the proposed solution are respectively about eight times and three times smaller than the Sidechains scheme. This paper also suggests a number of open issues and challenges faced by cross-metaverse interoperability that may inspire future research.

\end{abstract}


\section{Introduction}
\label{sec_intro}

The metaverse attracts significant attention in recent years. It is a computer-generated virtual world and can be intertwined with the physical world.
%
%
Users in the metaverse gain immersion during interacting with virtual objects or other users, which makes the metaverse a new paradigm of the Internet. 
To achieve this, the metaverse requires the integration of various technologies. These include but not limited to augmented reality (AR), digital twin (DT), blockchain, interactivity, game, artificial intelligence (AI), networking, and Internet of Things (IoTs) \cite{lee2021all}. For example, AR provides an immersive experience;  DT generates a mirror image of physical world entities; blockchains build metaverse  economic systems.

In metaverse, an avatar is the digital representation of a user within a virtual space. The avatar can be integrated with blockchain, to become the digital DNA of this user. In addition, a user may have digital assets in the metaverse, which can be cryptocurrencies, skins, or objects that feed from the physical world. Such assets become valuable and tradable by using blockchain. Further, blockchain, such as Ethereum, forms the fundamental economic system in the metaverse. On the other hand, non-fungible tokens (NFT) ensure that non-monetary digital assets owned by a user can't be copied or fabricated. 

To date, there are numerous metaverses have been built~{\color{black}\cite{Huang2022Economic}}. Examples include Roblox \cite{roblox} for productivity, Sandbox \cite{sandbox2023Sandbox} for virtual real estate, Axie Infinity \cite{axieinfinity} for gaming. Moreover, a number of top companies propose their own metaverse, such as Meta, Nvidia, and Google, to name a few~\cite{Huang2022Economic}. These metaverses usually have independent economic systems, i.e., they use different tokens or currencies, and  rules, i.e., representing avatars in 2D or 3D form. Hence, it is inconvenient for users to interchange between metaverses. This slackens the development of the metaverse society. Therefore, an efficient cross-metaverse strategy is desired to enable users to switch between metaverses with negligible effort. We call this \textit{cross-metaverse interoperability}. 




In a nutshell, cross-metaverse interoperability provides a seamless experience for users to interact with metaverses. We now use an example in the commercial metaverses to illustrate the key features of cross-metaverse interoperability. Consider a user purchased a digital asset, say an `axe', from metaverse \textit{Alpha}, the cross-metaverse interoperability ensures that this  asset has the same value and functions when the user using it in another metaverse \textit{Beta}. Note that, the metaverse \textit{Alpha} and \textit{Beta} may have different economic systems based on Bitcoin, Ethereum, or some type of centralized currency derived from offline central bank. Moreover, these metaverses may have different virtual environments, but the  cross-metaverse interoperability ensures that the digital `axe' is used to `cut trees' in all of them.

In order to achieve cross-metaverse interoperability, the following two basic components need to be interoperable between metaverses: 
%
\begin{itemize}
    \item {\em Identity}. 
    The identity defines the uniqueness of users and digital assets across metaverses. In addition, identities and their relationship are essential to connecting users with their actions and assets. 
    For cross-metaverse interoperability, identity is fundamental for building trustworthiness. This requires the standardization of identities among various types, such as users, assets, currencies, objects, and motions, in all metaverses. 
    It should be noted that one natural person may create multiple user identities in one or many metaverses. However, each of these identities needs to be treated independently, the same with multiple social accounts owned by the same person. 
    
    \item {\em Object}. This includes avatars, digital assets, and interactable entities in metaverses. Each object has its own properties, such as gender, material, rendering, and functionality. These properties can be fed from physical words, e.g., via DT and 3D scanning, or created by users out of thin air. For cross-metaverse interoperability, the objects in different metaverse with the same identity must have the same properties.

\end{itemize}




In this article, we first present a comprehensive review of existing technologies that enable cross-metaverse interoperability. 
We show that these technologies lack generality, making them hard to implement for interoperability between arbitrary centralized or decentralized metaverses.
We then propose a metaverse interoperability protocol named \textsf{MetaOpera}. It enables interoperability between any two metaverses regardless of their centralized or decentralized system models. By employing NFT technologies, all metaverse components are monetized as unique assets. Thus, they are exchangeable and satisfy the requirement of crossing the metaverses. As an exemplary concrete instantiation for \textsf{MetaOpera}, we describe two \textsf{MetaOpera} workflows in detail, which demonstrate how \textsf{MetaOpera} support metaverses interoperate with each other. 

To evaluate the validity of \textsf{MetaOpera}, we develop a proof-of-concept (PoC) implementation for \textsf{MetaOpera} metaverse and Sandbox metaverse. Without loss of generality, our \textsf{MetaOpera} metaverse is also applicable to any other metaverses. We evaluate the performance of \textsf{MetaOpera} and compare it with a cross-metaverse scheme based on Sidechains \cite{Peter2019Proof}. Experiment results show that the cross-metaverse proof of \textsf{MetaOpera} is roughly 1.7 KB, which is 8 times smaller than the Sidechains approach; and that the average time of cross-metaverse transactions is about 1.8 hours, which is 3 times smaller than the Sidechains approach. 
Lastly, we summarize open issues and challenges faced by cross-metaverse interoperability, which include economic systems, applications, physical/virtual interoperability, security, off-line trustworthiness, universal data structure, and policy.

The rest of this paper is organized as follows. Section \ref{sec:preliminaries} introduces the preliminaries and state-of-the-art cross-metaverse technologies. Section \ref{sec:MetaOpera} demonstrates the proposed \textsf{MetaOpera} protocol.
The performance of \textsf{MetaOpera} is evaluated in Section \ref{sec:performanceEvaluation}.
Section \ref{sec:OpenIssuesandChallenge} suggests open issues and challenges for 
cross-metaverse interoperability. Section \ref{sec:Conclusion} concludes this paper.

\section{Preliminaries and State-of-the-Art Cross-Metaverse Technologies}\label{sec:preliminaries}

According to the system model of metaverses, cross-metaverse interoperability is divided into two categories: (1) the interoperability between decentralized metaverses built at the top of blockchain technology, and (2) the interoperability between decentralized and centralized metaverses, where the centralized metaverses are built at the top of a centralized server. For the former, the underlying technology of interoperability is cross-chain. For the latter, the underlying technology of interoperability is on-chain and off-chain exchange. These two types of underlying technologies are described in detail below.
\begin{table*}[ht]
\caption{Cross-chain technologies}

\centering
\footnotesize
\renewcommand{\arraystretch}{1.5}
\begin{tabular}{|m{1.9cm}<{\centering}|m{3.3cm}|m{5cm}|m{1.cm}<{\centering}|m{1.cm}<{\centering}|m{1.5cm}<{\centering}|m{1.1cm}<{\centering}|}%
\hline
\textbf{Technical Scheme} & \textbf{Typical Project} & \textbf{Description}& \textbf{Security}& \textbf{Efficiency}& \textbf{Generality}& \textbf{Scalability} \\
\hline 
    Notary & Interledger \cite{thomas2015protocol}, Tokrex \cite{Tokrex}, Croda \cite{croda}, BTCB \cite{binance}, HBTC \cite{Hbtc}, tBTC \cite{tBTC}, ren \cite{ren}, DeCus \cite{decus}, Hop Exchange \cite{whinfrey2021hop}, Hyphen \cite{biconomy}, Degate Bridge \cite{degate}, Wanchain \cite{wanchain}, Fusion \cite{fusion} 
& A mutually trusted third party is used between different blockchains to act as a notary for cross-chain message verification and forwarding.&Low \cite{yin2022bool,xie2022zkbridge,bentov2019tesseract} & High &High& High\\
    \hline
     Hashed Time-Lock
 & WBTC \cite{wbtc}, Bridge \cite{cBridge}, Lightning Network\cite{lightningNetwork}, Zcash XCAT \cite{zcashXCAT}
& The asset receiver is forced to determine the collection and produce proof of collection to the payer within the cut-off time, or the asset will be returned via hash locks and blockchain "time" locks. The proof of receipt can be used by the payer to acquire assets of equal value on the recipient's blockchain or trigger other events.& Medium \cite{thyagarajan2022universal,tsabary2021mad} & High &Medium \cite{bentov2019tesseract}& Low\\
    \hline
  Sidechains/Relay
 &BTCRelay \cite{btcrelay}, RootStock \cite{rsk}, Plasma \cite{PLASMA}, Ronin \cite{Ronin}, Elements \cite{Elements}, LayerZero \cite{LayerZero}, Waterloo \cite{Waterloo}, Polygon \cite{Polygon}, 
 &Interoperation (transfer, communication, operation) of on-chain objects (assets, data, functions) through two-way anchoring and relay mechanisms.&High \cite{gavzi2019proof} & Medium \cite{gavzi2019proof,yin2021sidechains}&Medium \cite{zamyatin2019xclaim,yin2021sidechains}& High \cite{gavzi2019proof,yin2021sidechains}\\
    \hline
Relay chain
 &Polkadot \cite{Polkadot}, Cosmos \cite{kwon2019cosmos}
 & To achieve cross-chain object interoperation by the relay chain.& Medium & Low &Medium \cite{tian2021enabling}& High\\
    \hline
\end{tabular}
\label{Table:auction}
\end{table*}
\subsection{Cross-Chain Technologies}

To further understand cross-chain technologies, we review state-of-the-art cross-chain works. As described in table \ref{Table:auction}, there are four kinds of cross-chain technologies: Notary, Hash Time-Lock, Sidechains/Relay, and Relay chain. Notary scheme is the most widely used due to its high efficiency and convenient deployment. A typical example of a notary scheme is centralized exchange. All cross-chain assets in the exchange are maintained and managed by a trusted third party called notary. It is easy to see that if the notary is fail, cross-chain interoperability will be suspended. To avoid the single-point-failures, the notary consists of multiple parties based on their weights (i.e., coin) or reputation. However, the notary still suffers from security challenges such as external trust assumption. Indeed, Ronin bridge \cite{roninbridge}, a well-known notary scheme, lost 624 million USD due to malware attacks. The main reason behind this attack is that an attacker illicitly obtains the majority keys of the notary members through the malware.


To overcome the external trust assumption, the Hash time-lock scheme is proposed, which utilizes a hash function and time-lock features to achieve cross-chain interoperability. the security of the Hash time-lock scheme is based on cryptographic hardness assumptions. Consider two users: Alice with $x$ coins on chain $\mathcal{C}_1$ and Bob with $y$ coins on chain $\mathcal{C}_2$. They want to exchange their assets at the exchange rate of 1. First, Alice generates a hash value $p$ = $H$($p$) via a hash function $H(\cdot)$ and hereby uses it to create a transaction $tx_1$ that transfers $x$ coins to Bob's address on $\mathcal{C}_1$. That is, Bob can receive these $x$ coins from Alice if Bob knows the preimage $p$ corresponding to the $h$. Once the $tx_1$ is included on $\mathcal{C}_1$, Bob also uses the hash value $p$ to generate a transaction $tx_2$ that transfers $y$ coins to Alice's address on $\mathcal{C}_2$, and broadcasts it into $\mathcal{C}_2$ network. To gain these $y$ coins from Bob, Alice reveals the preimage $p$ in $tx_2$. Meanwhile, Bob also knows the $p$ via observing the $tx_2$ on $\mathcal{C}_2$. Subsequently, Bob publishes the obtained preimage $p$ in $tx_1$ and hereby receives these $x$ coins from Alice, completing the asset exchange. However, this scheme only supports monetary  exchange and thus has low scalability.

The sidechains/relay scheme supports the interoperability of multiple objects such as assets and other data, thus having high scalability. Sidechain is a blockchain that communicates with other blockchains via a two-way peg. In particular, the two-way peg is a mechanism that allows bidirectional communication between blockchains. An example of a two-way peg is simplified payment verification (SPV) in Bitcoin.
Specifically, to perceive events on a mainchain, all block headers of the mainchain are relayed into the sidechain pegged with the mainchain. The sidechain can determine whether an event has occurred on the mainchain by verifying the stored block headers of the mainchain and the Merkle tree verification path regarding the event. However, the sidechains/relay scheme can only be applied to specific blockchains, and thus has a low generality.

A relay chain is a third-party blockchain that other blockchains can connect to, which enables these blockchains to communicate with each other. The relay chain scheme has high scalability, and supports multiple objects interoperability between blockchains connected to the relay chain. However, as compared with other cross-chain schemes, the relay chain scheme has relatively low efficiency, especially in terms of cross-chain time.

\subsection{On-Chain and Off-Chain Technologies}

On-chain and off-chain technologies, often called oracle technologies, connect on-chain blockchain systems to off-chain systems, which further enables the interoperability of decentralized systems and centralized systems. Currently, oracles can be generally classified into two types: voting-based oracles and reputation-based oracles. The voting-based oracles are usually designed for  single centralized systems, which employ participants' weight to finalize interoperability outcomes. The widely deployed strategy of achieving these oracles includes stake-based oracles, multisignature-based oracles, schelling point-based oracles, and conventional oracles. However, these oracles are hard to verify the outcome integrity. On the other hand, reputation-based oracles are suitable for multiple centralized systems. Their aims are to authenticate the integrity of interoperability outcomes. These oracles utilize software (i.e., TLS protocol) or hardware (i.e., Intel Software Guard Extension) to generate a proof. The performance comparison of oracles is shown in Table \ref{Table:on-off-chain}.

\begin{table*}[ht]
\caption{Oracle technologies}

\centering
\footnotesize
\renewcommand{\arraystretch}{1.5}
\begin{tabular}{|m{3cm}<{\centering}|m{2.6cm}|m{5.2cm}|m{1.3cm}<{\centering}|m{1.3cm}<{\centering}|m{1.9cm}<{\centering}|}%
\hline
\textbf{Technical Scheme} & \textbf{Specific Type} & \textbf{Description}& \textbf{Security}& \textbf{Integrity}& \textbf{Confidentiality} \\
\hline 
 \multirow{4}{*}{Voting-based Oracles}   
& Stake-based \cite{merlini2019public, adler2018astraea, ryuuji2018Shintaku, nelaturu2020public, peterson2015augur, tellor2021Tellor, dIA2021Decentralized, paul2021Truthcoin} & This scheme employs the stake held by participants for finalizing the outcome. Participants are rewarded if the outcome is matched and penalized otherwise. & High & Low & Low \\
    \cline{2-6}
      & Multi-signature based \cite{orisi2014Orisi, gnosis2017Gnosis, delphi2017Delphi, moudoud2019iot, dOS2019DOS}
& Finalizing the outcome is determined collectively by multi-participants via making a signature on the outcome. & Medium & Medium & Low \\
    \cline{2-6} 
 & Schelling point based \cite{vitalik2014SchellingCoin, roman2017Oracul, MarkerDAO2014The}  & According to the median value, the outcome is derivated from a group of data providers' answers. & Low  & Low & Low  \\
    \cline{2-6}
 & Conventional \cite{IOTA2021Introducing, eskandari2017feasibility, zhang2020industrial, Thomas2020ternity}
 & The outcome is directly from the raw answer of the data provider.& Low & Low & Low \\
    \hline
 \multirow{3}{*}{Reputation-based Oracles} & Software-based Proof \cite{zhang2020deco, guarnizo2019pdfs, tlsnotary2023TLSnotary, bridge2021oracle, liang2021polkaoracel}
 & Utilizing the TLS/SSL protocols, the scheme generates a data authenticate proof which is used to verify the integrity of outcomes.& Medium & Medium & Medium \\
    \cline{2-6}
 & Hardware-based Proof \cite{zhang2016town, schaad2019integration, coldcard2021coldcard, ellis2017chainlink} & The scheme provides the integrity and conidentiality for the outcome based on the Software Guard Extension (SGX) technology in the Inter CPUs. & Medium & High & High \\
    \cline{2-6}
 & Proofless \cite{wang2019novel, al2019decentralized, fujihara2019proposing, de2017witnet}
 & The scheme retrieves directly from data sources without providing authenticated proof of the outcome. & Low & Low & Low \\
    \hline
\end{tabular}
\label{Table:on-off-chain}
\end{table*}

\subsection{Cross-Metaverse Technologies}

In terms of cross-metaverse interoperability,
industrials have proposed prototypes.
For example, \textit{STYLE} protocol \cite{styleprotocol2023}, which is a virtual asset infrastructure across-metaverse, has been proposed in 2022 and will be implemented this year. The goal of \textit{STYLE} is to enable assets to exchange across the metaverses. To achieve this, two key features: usability and visualization for assets, are introduced into this protocol. Cross-metaverse avatar scheme \cite{crossmetaverse2023} aims to enable avatar interoperability across-metaverse. Same with in the real world to interact and communicate, avatars, a digital representation in metaverse, also allow people to feel and travel in multiple metaverses. Similarly, the scheme will be launched in 2023. In academics, Huang et al. \cite{Huang2022Economic} presented a cross-chain ecosystem for metaverse. As an example, this work depicts assets interoperability between Sandbox’s and Axie Infinity’s metaverses. However, this study focuses only on decentralized metaverses. Chen et al. \cite{chen2022cross} proposed a cross-platform metaverse data management system. The system designs plug-ins for different metaverse/games, enabling the profile and space share cross-platform metaverse.

\section{\textsf{MetaOpera} Protocol}\label{sec:MetaOpera}
\begin{figure*}[t]
    \centering
\includegraphics[width=0.95\textwidth]{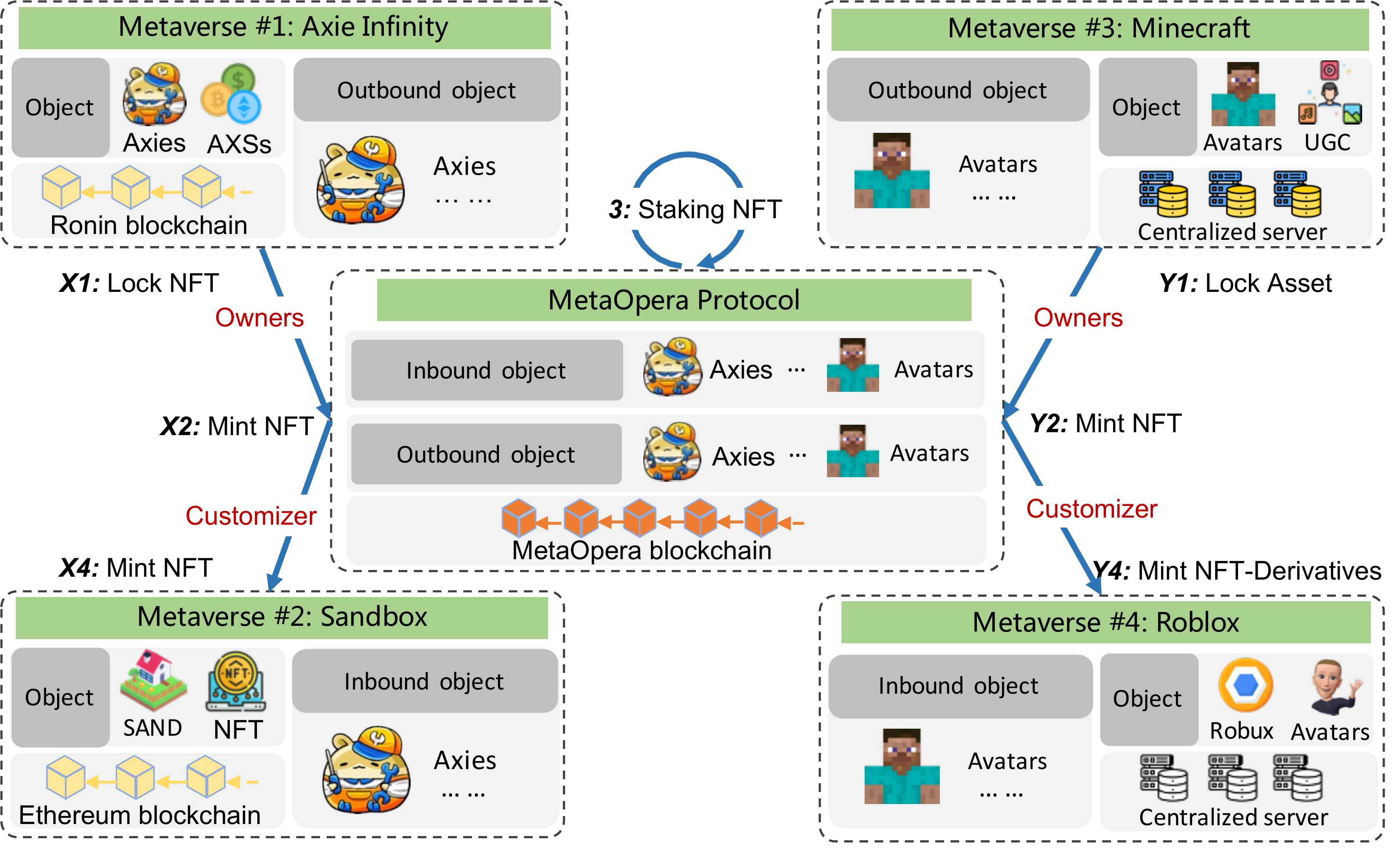}
    \caption{The design of \textsf{MetaOpera} illustrated with the example of four metaverse interoperability. The blue arrows denote the direction of cross-metaverse object transfer. For clarity we only show two workflows of \textsf{MetaOpera} and the opposite direction is symmetric.}
    \label{fig:MetaOperaProtocol}
\end{figure*}
In this section, we present a novel cross-metaverse interoperability protocol: \textsf{MetaOpera}. We first give an overview of \textsf{MetaOpera}, and then present its core components and workflow.

\subsection{Overview}

At a high level, \textsf{MetaOpera} is a cross-metaverse interoperability infrastructure, which enables interoperability between any two metaverses. \textsf{MetaOpera} implements NFT technologies to monetize all metaverse components as unique assets. Therefore, arbitrary metaverse users and objects are able to interoperate with each other. Its key idea is shown in Figure \ref{fig:MetaOperaProtocol}. Briefly, the user of the metaverse connected to \textsf{MetaOpera} first locks its asset, and then mints the corresponding asset, in the form of NFT, in the \textsf{MetaOpera}. Next, the user of the \textsf{MetaOpera} stakes its NFT and then mints the corresponding NFT in another metaverse it wishes, completing metaverse object interoperability. 

\subsection{\textsf{MetaOpera} Core Components}


MaoetaOpera is essentially a relay metaverse built at the top of \textsf{MetaOpera} blockchain, which is used to connect another metaverse. Upon connecting to the \textsf{MetaOpera}, any metaverse can communicate with each other. To achieve a variety of metaverse interoperability, \textsf{MetaOpera} has two key underlying technologies: cross-chain technology and on-chain and off-chain technology. The former can support \textsf{MetaOpera} to interoperate with the decentralized metaverse based on blockchain. The latter can achieve \textsf{MetaOpera} to interoperate with the centralized metaverse built at top of centralized servers. these underlying technologies adopted in the \textsf{MetaOpera} are resilient. To reach high efficiency across decentralized metaverse, for instance, the cross-chain scheme may be a notary technology as described in Table \ref{Table:auction}.



There are five components in \textsf{MetaOpera} listed as follows:

\begin{itemize}
    \item Metaverse. A metaverse is a virtual world. Metaverses are classified into two categories: decentralized metaverse (DM) and centralized metaverse (CM) based on underlying technologies. Here, DM is built at the top of decentralized blockchain technology. while CM is built at the top of centralized serves.
    \item Object. An object is a virtual digital item in the metaverse, such as cryptocurrency, avatar, skin, pet, fashion, etc. These objects can be transferred, sold, rented, and staked across metaverses. Each metaverse includes \textit{outbound objects} and  \textit{inbound objects}. An object is referred to as an outbound object if it had been locked in the metaverse it belongs; subsequently, a corresponding object will be minted in another metaverse. On the contrary, an inbound object denotes the object moved from another metaverse.
    \item Owner. An owner represents a metaverse user, who holds some objects and wishes to transfer them from one metaverse to another metaverse. 
    \item Customizer. A customizer is a \textsf{MetaOpera} worker who assists in object transformation from \textsf{MetaOpera} to another metaverse in exchange for a reward. For example, the 2D object of \textsf{MetaOpera} can be tailored into the 3D object residing on a metaverse through a customizer.
    \item NFT. NFT (non-fungible token) is a unique and exchangeable digital token. NFT has the characteristics of rareness, indivisibility, and uniqueness. Thus, it can denote the ownership of an object from metaverse, specifically for the centralized metaverse.
\end{itemize}

\subsection{\textsf{MetaOpera} Workflow}

In this subsection, we demonstrate the interoperability workflow of the \textsf{MetaOpera} by depicting two across-metaverse interoperability scenarios, see in Figure \ref{fig:MetaOperaProtocol}. 
The first workflow is DM to DM, e.g., Axie Infinity metaverse \cite{axieinfinity} to \textsf{MetaOpera} to Sandbox metaverse \cite{sandbox2023Sandbox}. The other workflow is CM to CM, e.g., Minecraft metaverse \cite{minecraft} to \textsf{MetaOpera} to Roblox metaverse \cite{roblox}. It is worth noting that \textsf{MetaOpera} can support any two metaverses to interoperate regardless of the CM or DM models of the metaverse.

\subsubsection{DM to \textsf{MetaOpera} to DM}

   As depicted in Figure\ref{fig:MetaOperaProtocol}, two decentralized metaverses (DM), Axie Infinity and Sandbox, can communicate with each other through \textsf{MetaOpera}. Specifically speaking, the owner of Axie Infinity wishes to transfer assets into Sandbox. First, the owner issues a transaction $tx_{x1}$ that locks its asset \textit{Axies}. Once it has been locked, \textit{Axies} will become an outbound object in Axie Infinity. Meanwhile, the proof attesting to the validity of $tx_{x1}$ is generated by a committee of \textsf{MetaOpera}. Here the committee is a notary scheme described in Figure \ref{Table:auction}, The members of the committee select from the maintainer of \textsf{MetaOpera} according to their weights (i.e., coins). If the above $tx_{x1}$ and its proof are considered valid, \textsf{MetaOpera} will mint an NFT corresponding to \textit{Axies}, The minting action is achieved by a mint transaction $tx_{x2}$ generated by the owner. Until now, the owner of Axie Infinity has transformed its \textit{Axies} into the NFT of \textsf{MetaOpera}.
    
    Further, the owner sequentially transfers the NFT corresponding to \textit{Axies} into Sandbox. This process is similar to the transfer from Axie Infinity to \textsf{MetaOpera}. The main difference is that a customizer may participate in the transformation of object format, such as casting 2D NFT into 3D NFT, The reason behind the difference is the heterogeneity of metaverse model so that an object of a metaverse is incompatible with another metaverse version. The owner first stakes its NFT corresponding to \textit{Axies} in \textsf{MetaOpera} by issuing a staking transaction \textit{$tx_{x3}$}. According to the NFT format, a customizer tailors it into another NFT format supported by Sandbox. Subsequently, proof testifying that \textit{$tx_{x3}$} is valid is produced by the committee of \textsf{MetaOpera}. Finally, Sandbox mints corresponding NFT to the owner based on the proof and NFT format cast by the customizer, completing the interoperability for Axie Infinity and Sandbox.

\subsubsection{CM to \textsf{MetaOpera} to CM}
    
    Upon connecting to the \textsf{MetaOpera}, any two centralized metaverse (CM) can interoperate with each other. Figure \ref{fig:MetaOperaProtocol} shows such an example that an avatar of Minecraft travels to Roblox. The owner of Minecraft generates a lock transaction \textit{$tx_{y1}$} to lock its avatar. A committee of \textsf{MetaOpera} observes the Minecraft state. Here the committee is a multi-signature oracle described in Table \ref{Table:on-off-chain} and consists of \textsf{MetaOpera} users. If \textit{$tx_{y1}$} is confirmed in Minecraft. The committee will generate a proof for \textit{$tx_{y1}$} to convince all \textsf{MetaOpera} maintainers that the avatar has been locked successfully. Once the proof is generated, the owner issues a mint transaction \textit{$tx_{y2}$} to mint an NFT corresponding to the avatar into its \textsf{MetaOpera} address. The action of minting the NFT will be implemented if the \textit{$tx_{y2}$} and proof are considered valid. This means that the avatar has transferred from Minecraft to \textsf{MetaOpera}.
    
    Next, the owner further transfers the avatar from \textsf{MetaOpera} to Roblox. The owner first generates a staking transaction \textit{$tx_{y3}$} to lock the NFT above. Similarly, the proof for \textit{$tx_{y3}$} is also generated. A customizer issues a mint transaction \textit{$tx_{y4}$} in Roblox to cast an NFT-derivative corresponding to the locked NFT. Where NFT-derivative is not NFT but an object compatible with Roblox, since Roblox does not support NFT. To keep the format persistence of the NFT-derivative and the Minecraft avatar, casting NFT-Derivative is essential. This is done by the customizer. The above casting action will be executed eventually if the \textit{$tx_{y4}$} and proof are considered valid. Until now, the avatar of Minecraft has been transferred into Roblox, completing the interoperability for Minecraft and Roblox. 

\subsubsection{DM/CM to \textsf{MetaOpera} to CM/DM}

As discussed earlier, \textsf{MetaOpera} is a decentralized metaverse built at the top of the blockchain. 
Objects are able to interoperate from DM or CM to \textsf{MetaOpera} and then interoperate to DM or CM again. 
Alternatively, \textsf{MetaOpera} can support objects interoperating from DM to CM or CM to DM.


\section{Performance Evaluation}\label{sec:performanceEvaluation}

In this section, we implement the proposed \textsf{MetaOpera} protocol and evaluate its performance. We first measure the cross-metaverse proof size and the average time of cross-metaverse transactions. We then compare them with a state-of-the-art cross-metaverse scheme based on Sidechains \cite{Peter2019Proof}.

\textbf{Implemention Setting.} To simulate the performance of \textsf{Metaopera} prorocol from Section \ref{sec:MetaOpera}, we make a proof-of-concept (PoC) implementation for \textsf{Metaopera} and Sandbox. Without loss of generality, the PoC implementation is also applicable to both \textsf{Metaopera} and anyone metaverse connected to \textsf{Metaopera}. In our PoC implementation, the blockchain of \textsf{Metaopera} is instantiated as Cardano \cite{aggelosre2017Ouroboros}. While the blockchain of Sandbox is based on Ethereum \cite{buterin2014next}, which follows the Sandbox implementations of \cite{sandbox2023Sandbox}). The PoC implementations for \textsf{MetaOpera} and Sandbox are executed in a personal platform installed with Windows 11 operating system and equipped with Intel(R) Core(TM) i7-12700KF CPU 3.60 GHz 32.00GB RAM. We carry out our \textsf{MetaOpera} blockchain in standard C language. To achieve committee members' vote, the multi-signature scheme \cite{boldyreva2003Threshold}, a cryptographic primitive, is employed in our PoC implementations. Here the size of public key $|vk_i|$ = 272 bits, the size of signature $|\sigma_i|$ = 528 bits, the size of proof-of-possession $|POP|$ = 528 bits. A 256-bit hash function $H(\cdot)$ is also applied in the PoC implementations, meaning that $|H(\cdot)|$ = 256 bits. Following Cardano implementation, we denote by $k$ a common prefix parameter. To form a decentralized \textsf{Metaopera} committee, the chain quality, a blockchain fundamental security property, is adopted in the committee selection. In short, the committee members are selected from nodes that have generated any $k$ consecutive blocks of Cardano. Moreover, we set that the transaction size $|tx|$ = 250 bytes, block height size $|h|$ = 32 bytes, and the random number size $|r|$ = 32 bytes, these parameters are important components of the above proof. On the other hand, we use $k^{\prime}$ to indicate a common prefix parameter of Ethereum. In addition, both the block-generating times of Cardano and Ethereum are set to 5 seconds.

\begin{figure}[ht]
	\centering
	\includegraphics[width=0.400\textwidth]{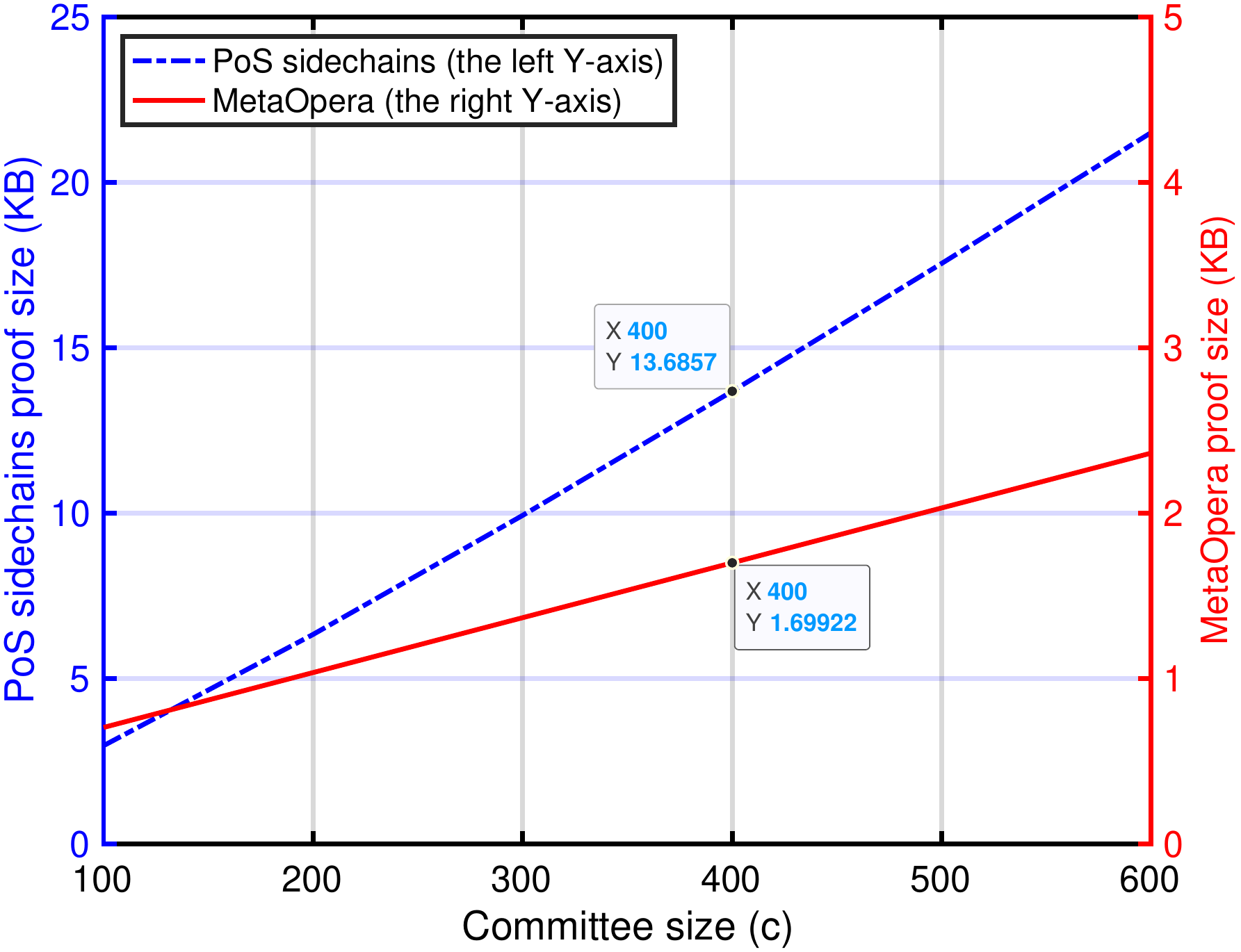}
	\caption{The proof size at different committee sizes.}
	\label{CommitteeProof}
\end{figure}

\textbf{Implementations and Evaluations.} In our \textsf{Metaopera}, the proof attesting to the validity of cross-metaverse transactions is generated by a committee of \textsf{Metaopera}. Its size affects the efficiency of cross-metaverse transactions, For example, The larger the proof size is, the longer its verification time will be. Thus, we evaluate it. A proof is composed of public keys, signatures, and hash values. Based on the principle of the adopted multi-signature scheme, the proof can be further improved by the optimized tips used in \cite{Peter2019Proof}. Formally speaking, the improved proof can be denoted as
$|tx|+|h|+|r|+0.1\cdot k \cdot |vk_i|+ |\sigma_i|$. As depicted in Figure \ref{CommitteeProof}, the proof size of \textsf{Metaopera} increases linearly in the size of committee $\textit{c}$. This is because the proof size increases as the number of committee members increases. More precisely, in the case of the committee size of 400, the proof size is roughly 1.7 KB, which is 8 times smaller than the proof size of PoS sidechains \cite{Peter2019Proof}. 

In our PoC implementation, cross-metaverse transactions consist of two transactions: \textit{tx} included in \textsf{Metaopera} and  \textit{$tx^{\prime}$} included in Sandbox. To verify the validity of \textit{$tx$}, the proof is essential. Therefore, the time of cross-metaverse transactions includes both the confirmed time of \textit{tx} and \textit{$tx^{\prime}$}, as well as the proof generating time. Let \textit{k} and \textit{$k^{\prime}$} be 400 and 500 respectively. We have that the confirmed time of \textit{tx} is roughly 5\textit{k} = 5 $\times$ 400 = 33.33 minutes and the confirmed time of \textit{$k^{\prime}$} is roughly 5\textit{$k^{\prime}$} = 5 $\times$ 500 = 41.67 minutes. In the case of a committee size of 400, we measure the time \textit{t} that these committee members collectively generate the proofs for 200 different cross-metaverse transactions. Measurement results show that the average time for generating a proof is roughly 1.5 minutes. To this end, the average time of cross-metaverse transactions is roughly 76.5 minutes, as described in Table \ref{Table:TimeCrossMetaverse}. As a comparison, the average time of transactions of Sidechains \cite{Peter2019Proof} is 5.251 hours, which is 3 times more than our \textsf{MetaOpera}.

\begin{table*}[ht]
\caption{The average time of cross-metaverse transactions.}
\centering
\footnotesize
\renewcommand{\arraystretch}{1.5}
\renewcommand{\arraystretch}{1.5}
\begin{tabular}{|m{5.5cm}<{\centering}|m{2.2cm}<{\centering}|m{2.4cm}<{\centering}|m{2.4cm}<{\centering}|m{2.6cm}<{\centering}|}%
\hline
\textbf{Time of cross-metaverse transaction} & \textbf{\textsf{MetaOpera} time} & \textbf{\textsf{MetaOpera} time/h} & \textbf{Sidechains \cite{Peter2019Proof} time} & \textbf{Sidechains \cite{Peter2019Proof} time/h} \\ 
\hline
\textbf{Confirm Time of \textit{tx} in \textsf{MetaOpera}} & \textit{2k} & 1.112 & \multirow{2}{*}{\textit{10k}} & \multirow{2}{*}{5.556} \\ 
\cline{1-3}
\textbf{Proof Generating Time} & \textit{t} & 0.025 & & \\
\hline
\textbf{Confirm Time \textit{$tx^{\prime}$} in Sandbox} & \textit{$k^{\prime}$} & 0.695 & \textit{$k^{\prime}$} & 0.695 \\
\hline
\textbf{Average time of cross-metaverse transaction} & \textit{2k} + \textit{t} + \textit{$k^{\prime}$} & 1.832 & \textit{10k} + \textit{$k^{\prime}$} & 6.251 \\ 
\hline
\end{tabular}
\label{Table:TimeCrossMetaverse}
\end{table*}

\section{Open Issues and Challenges}\label{sec:OpenIssuesandChallenge}
Although we have proposed the \textsf{MetaOpera} protocol, cross-metaverse interoperability is still in its initial stage from the perspective of either industry or academia. In this section, we summarize open issues and challenges in cross-metaverse interoperability and suggest potential future research directions. 

\subsection{Economic System}

To date, most metaverse builds their economic system using smart contracts on the Ethereum blockchain. This raises the risk that a single security hazard in Ethereum will affect a majority of metaverses. To this end, diversified economic systems are desired in co-existing metaverses, to enhance their robustness against single-system crashes. This also requires cross-metaverse interoperability of these systems. 


  

\subsection{Applications}
  At the current stage of metaverse development, almost all the applications 
  provided by the blockchain-based metaverses are the circulation of cryptocurrencies and NFT. However, the metaverse has more potential applications other than this. For example, the immersive experience can be used for a deep interaction between humans and multiple physical or virtual environments. This needs further development of cross-metaverse interoperability technologies to boost the application scenarios of the metaverse.  

\subsection{Physical/Virtual Interoperability }
Most research and implementation of metaverse focus on mapping physical objects to the virtual metaverse. There are a number of technologies that can be used, such as 3D scanning. However, how the actions in the virtual world reflect in the physical world is also a key concern. This is important for production metaverse applications such as surgery and driving. Such interoperability requires trustworthiness, low latency, and accuracy of the reflection of the action. 


%
 

\subsection{Security}
The security issues in cross-metaverse interoperability concern with authentication and permissions for assets and data exchanges. In particular, when an asset or object data is crossing metaverses, both metaverses need to confirm their ownership. Moreover, these metaverses need to ensure the asset or data is only valid in only one metaverse. For the objects existing in a metaverse, permission is important to specify who and what actions can be interacting with the objects. 

%

\subsection{Off-line Trustworthiness}
Even though the Internet is universal existence in the physical world, it is inevitable that some locations still have no Internet connection, such as in a desert. After an object lost connection with the metaverse, its properties may be changed before back online. In such a situation, cross-metaverse interoperability needs to ensure the trustworthiness of the changes from a tiny centralized server, i.e., an offline data collector, to the centralized or decentralized online metaverse. 



%
%

\subsection{Universal Data Structure}
Physical objects have common properties, such as shape, weight, and color. However, some of their properties are unique and independent. For example, a cup has the capacity property, but a pencil does not. Hence, cross-metaverse interoperability needs to design a universal data structure to describe the physical objects in the metaverse. In addition, this data structure needs to be a standard for all metaverses to enable interoperability. 



\subsection{Universal Policy}

The challenges of the universal policy include two perspectives: governance and metaverse platform. In terms of governance policy, a universal law or rule is required to restrain criminal behavior. This is because metaverses have independent rules, and their evaluation criteria for improper behavior are different. Hence, a universal governance policy is vital to prevent cross-metaverse crimes.  On the other hand, existing metaverse platforms usually have different user policies. Hence, users are hard to control the privacies, copyrights, and exchange of their assets in multiple metaverses. Therefore, a universal metaverse platform policy is also critical for  cross-metaverse interoperability. 




\section{Conclusion}\label{sec:Conclusion}
Cross-metaverse interoperability is a fundamental requirement when users move their digital assets between diverse metaverses. This paper first introduces the preliminaries and basics of interoperability of cross-metaverse. It then proposes the \textsf{MetaOpera} protocol for cross-metaverse interoperability and evaluated its performance. This paper also discusses the challenges and open issues of cross-metaverse interoperability. Furthermore, we hope this article is capable to inspire researchers, engineers, and educators to explore more cooperative metaverse applications to establish a better metaverse society.
 
\bibliographystyle{IEEEtran}
\bibliography{reference}

\begin{IEEEbiographynophoto}{Taotao~Li}
received the Ph.D. degree in cyber security from the Institute of Information Engineering, Chinese Academy of Sciences and University of Chinese Academy of Sciences, China, in 2022. He is currently a postdoc with the School of Software Engineering, Sun Yat-Sen University, Zhuhai, China. His main research interests include blockchain, Web3, and applied cryptography.
\end{IEEEbiographynophoto}

\begin{IEEEbiographynophoto}{Changlin~Yang}
[M'17] obtained his PhD degree from the University of Wollongong in 2015. He was a Senior Wireless Engineer at Huawei Australia from 2015 to 2017, a lecturer at the School of Computer Science, Zhongyuan University of Technology, China from 2017 to 2018 and 2021 to 2022, and a postdoctoral researcher at the Department of Electrical Engineering, Columbia University, USA from 2019 to 2021. He is now a research fellow with the School of Software Engineering, Sun Yat-Sen University. His research interests include coverage problems in Internet of Things, blockchain scalability and trustworthiness metaverses. 
\end{IEEEbiographynophoto}

\begin{IEEEbiographynophoto}{Qinglin~Yang}[M'21]  received his B.S. degree from the School of Geographical Sciences in 2014 and received his M.S. degree of computing mechanism from College of Civil Engineering, Kunming University of Science and Technology in 2017, and a Ph.D. degree from the University of Aizu, Japan in computer science and engineering and 2021. He is a research fellow at School of Intelligent System Engineering, Sun Yat-sen University, China. His research interests include edge computing, federated learning, and Web3.
\end{IEEEbiographynophoto}

\begin{IEEEbiographynophoto}{Siqi~Zhou}
is now pursuing a Ph.D. degree in mathematics from the School of Mathematical Sciences, Shanghai Jiao Tong University, China. 
Her main research interests include quantum information, quantum algorithm, and blockchain.
\end{IEEEbiographynophoto}

\begin{IEEEbiographynophoto}{Huawei~Huang}
[SM'22] (corresponding author, huanghw28@mail.sysu. edu.cn) is an Associate Professor at Sun Yat-Sen University. He received his Ph.D. degree from the University of Aizu (Japan) in 2016. He has served as a research fellow of JSPS, and a program-specific Assistant Professor at Kyoto University, Japan. His research interests include blockchain and distributed computing. He has served as a lead guest editor of blockchain special issues at IEEE JSAC and IEEE OJ-CS. He also served as a TPC chair for multiple blockchain conferences and workshops.
 
\end{IEEEbiographynophoto}

\begin{IEEEbiographynophoto}{Zibin~Zheng} (~Fellow, IEEE) is a Professor and Deputy Dean of the School of Software Engineering, Sun Yat-sen University, China. He published over 200 international journal and conference papers. According to Google Scholar, his papers have more than 26,000 citations. His research interests include blockchain, software engineering, and services computing. He was a recipient of several awards, including the IEEE TCSVC Rising Star Award, IEEE Open Software Award, Top 50 Influential Papers in Blockchain, the ACM SIGSOFT Distinguished Paper Award of ICSE, and the Best Student Paper Award at ICWS.
\end{IEEEbiographynophoto}

\end{document}